\documentclass{emulateapj}
\usepackage{amsmath}
\usepackage{float}

\slugcomment{Submitted to The Astrophysical Journal}

\begin{document}

\title{Carbon Shell or Core Ignitions in White Dwarfs Accreting from Helium Stars}

\author{Jared Brooks\altaffilmark{1}, Lars Bildsten\altaffilmark{1,2}, Josiah Schwab\altaffilmark{3,4}, Bill Paxton\altaffilmark{2}}

\altaffiltext{1}{Department of Physics, University of California, Santa Barbara, CA 93106}
\altaffiltext{2}{Kavli Institute for Theoretical Physics, Santa Barbara, CA 93106}
\altaffiltext{3}{Physics Department, University of California, Berkeley, CA 94720, USA}
\altaffiltext{4}{Astronomy Department and Theoretical Astrophysics Center, University of California, Berkeley, CA 94720, USA}

\begin{abstract}

White dwarfs accreting from helium stars can stably burn at the accreted rate and avoid the challenge of mass loss associated with unstable Helium burning that is a concern for many Type Ia supernovae scenarios. 
We study binaries with helium stars of mass $1.25 M_\odot\le M_{\rm{He}} \le 1.8 M_\odot$, which have lost their hydrogen rich envelopes in an earlier common envelope event and now orbit with periods ($P_{\rm orb}$) of several hours with non-rotating $0.84$ and $1.0 M_\odot$ C/O WDs.  
The helium stars fill their Roche lobes (RLs) after exhaustion of central helium and donate helium on their thermal timescales (${\sim}10^5$yr). 
As shown by others, these mass transfer rates coincide with the steady helium burning range for WDs, and grow the WD core up to near the Chandrasekhar mass ($M_{\rm Ch}$) and a core carbon ignition. 
We show here, however, that many of these scenarios lead to an ignition of hot carbon ashes near the outer edge of the WD and an inward going carbon flame that does not cause an explosive outcome. 
For $P_{\rm orb} = 3$ hours, $1.0 M_\odot$ C/O WDs with donor masses $M_{\rm He}\gtrsim1.8 M_\odot$ experience a shell carbon ignition, while $M_{\rm He}\lesssim1.3 M_\odot$ will fall below the steady helium burning range and undergo helium flashes before reaching core C ignition. 
Those with $1.3 M_\odot \lesssim M_{\rm He} \lesssim 1.7 M_\odot$ will experience a core C ignition. 
We also calculate the retention fraction of accreted helium when the accretion rate leads to recurrent weak helium flashes.

\end{abstract}


\section{Introduction}\label{sec:intro}

The possible progenitor systems for SNe Ia presently fall into two categories: the single degenerate and double degenerate scenarios, each with theoretical challenges.
The double degenerate scenario, characterized by the merger of two WDs that unstably ignites degenerate carbon, is challenged because an off-center ignition of carbon likely converts C/O WDs to O/Ne WDs via an inward-propagating carbon flame \citep{Nomoto1985, Saio1985, Woosley1986, Saio1998, Shen2012}.
Recent 3D simulations of these mergers found a prompt detonation can be triggered during the merger process when the mass ratio is close to unity \citep{Pakmor2010, Pakmor2011, Pakmor2012, Ruiter2012}.
The single degenerate scenario, characterized by stable accretion onto WDs until they grow to the Chandrasekhar mass ($M_{\rm Ch}$), is challenged by many theoretical and observational issues, including hydrogen flashes, or flashes of the helium built up via steady hydrogen burning, which remove mass, possibly preventing efficient growth of the core \citep{Iben1989, Nomoto2007, Wolf2013}.
Included in the single degenerate scenario, however, are systems with helium star donors of mass $M_{\rm He}\approx 1.2-1.8 M_\odot$ that donate He-rich matter to WDs at $\dot{M}>10^{-6}M_\odot$yr$^{-1}$. 
This avoids hydrogen flashes, and, given a certain range of helium star masses, can allow for steady helium burning on the surface of the WD at the rate that it is accreted \citep{Yoon2003, Shen2007, Piersanti2014a} leading to steady growth of the WD core and a possible core carbon ignition.
However, at these large $\dot{M}$s, another possible outcome is a shell ignition of carbon that will non-explosively convert a C/O WD to an O/Ne WD, leading eventually to an accretion induced collapse (AIC) rather than a SN Ia \citep{Nomoto1985, Saio1985, Saio1998}.

\cite{Wang2009} explore this channel using an optically thick wind model \citep{Hachisu1996}, instead of solving the stellar structure equations of the accreting WDs. 
They find the area in the initial orbital period $-M_{\rm He}$ plane where binary systems with these initial parameters will end in core ignitions.
Assuming that the core ignitions lead to Type Ia SNe, they perform binary population synthesis (BPS) studies and calculate the birthrate of SNe Ia. 
By not solving the stellar structure equations of the WDs, however, they do not take into account the possibility of non-explosive shell carbon ignition. 
As we show, the effect of including this possibility shrinks the area in the $\log P_{\rm orb,0}-M_{\rm He}$ plane that lead to SNe Ia, and lowers the estimated SNe Ia rate through this channel. 

In \S \ref{sec:steady} we calculate the boundaries for steady helium burning accretion rates using \texttt{MESA} \citep{Paxton2011, Paxton2013, Paxton2015a}, and explain the behavior of the models above and below the steady helium burning range. 
Then in \S \ref{sec:core} we explore the behavior of the core as it grows to core ignition near $M_{\rm Ch}$ and explain the ``race'' to ignition between carbon the core and in the shell. 
We include in \S \ref{sec:dev} a discussion on the effects of neutrino cooling the cores of the WDs, discuss the possible observables from these systems in \S \ref{sec:hr}, and explore a low mass binary case that leads to a double degenerate scenario in \S \ref{sec:lowmass}.
We conclude in \S \ref{sec:conc}.

\section{Steady Helium Burning}\label{sec:steady}

When binary systems have high mass helium star donors ($1.3 M_\odot\le M_{\rm{He}} \le 1.8 M_\odot$) the mass transfer is driven by the expansion of the donor as core helium is exhausted and the star leaves the He-main sequence.
This allows for mass transfer rates that cross the regime for steady He burning \citep{Piersanti2014a}.

\subsection{Calculation of Steady He burning boundaries}\label{sec:bound}

\begin{figure}[H]
  \centering
  \includegraphics[width = \columnwidth]{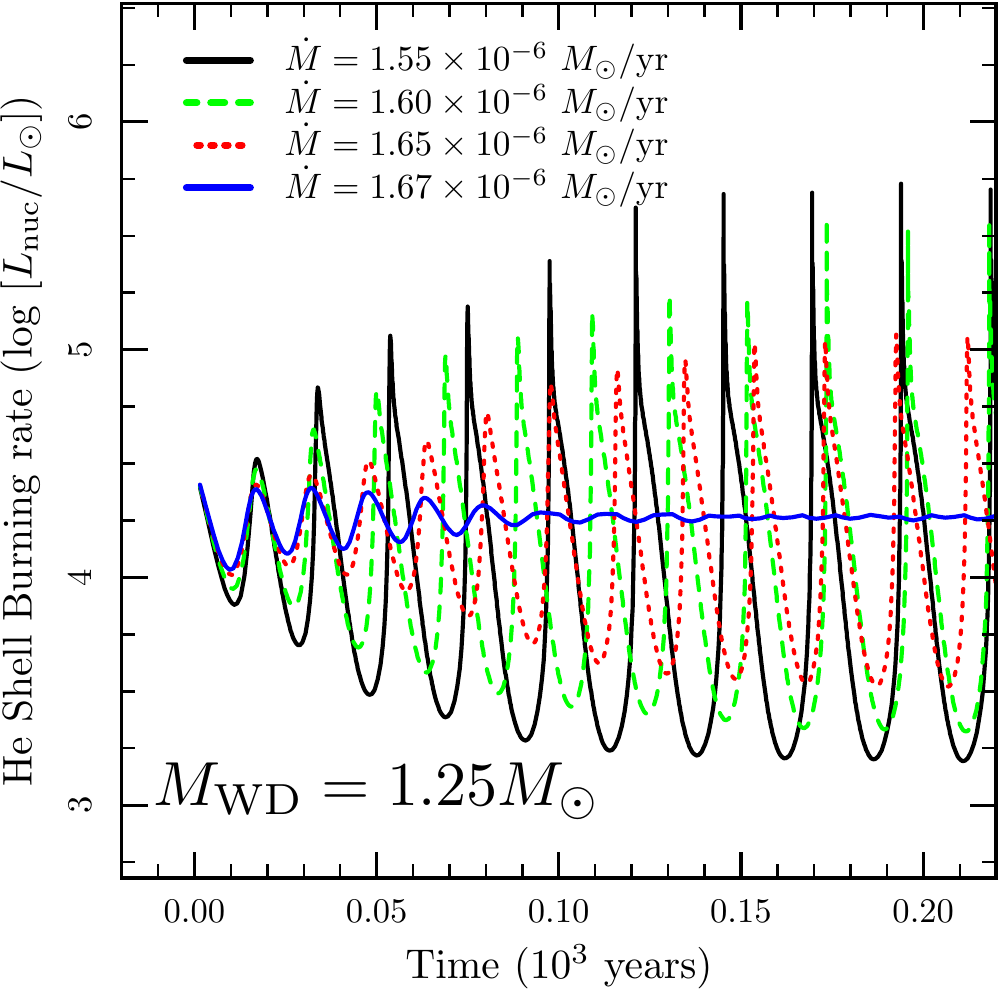}
  \caption{\footnotesize Nuclear burning rates for a narrow set of accretion rates near the lower stability boundary for a $1.25 M_\odot$ WD. 
  Starting from a stable burning model at $\dot{M}_{\rm WD}=3.5\times10^{-6} M_\odot\text{ yr}^{-1}$, the accretion rate is lowered to shown values. 
  Above a certain accretion rate the oscillations in the burning rate are effectively damped.}
  \label{fig:6}
\end{figure}

\begin{figure}[H]
  \centering
  \includegraphics[width = \columnwidth]{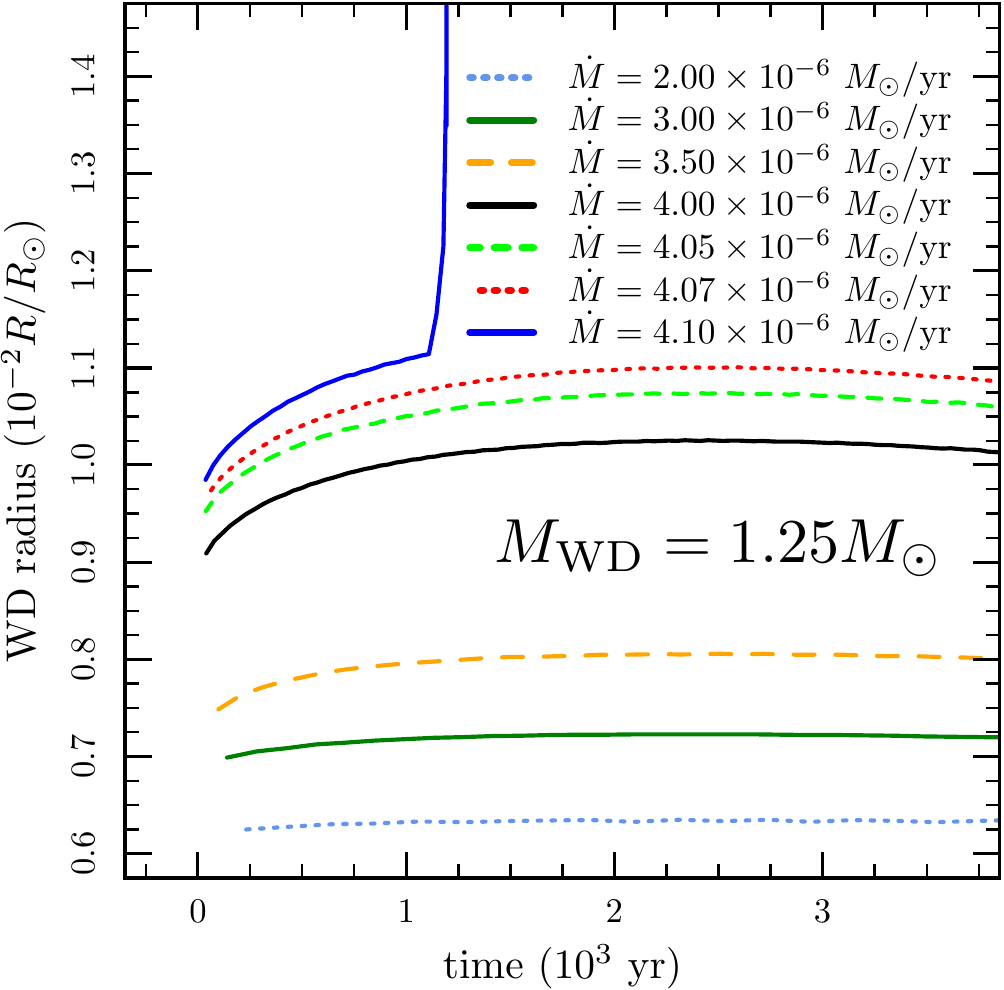}
  \caption{\footnotesize Starting from $\dot{M}_{\rm WD}=3.5\times10^{-6} M_\odot\text{ yr}^{-1}$ on a $1.25 M_\odot$ WD, the accretion rate is raised to the shown value. 
  Above a certain accretion rate, the radius rapidly expands as the WD enters the RG regime.}
  \label{fig:7}
\end{figure}

The steady burning boundaries are calculated using \texttt{MESA} (r7624) by taking a set of WD models that span the mass range [0.8,1.397] $M_\odot$ and running them each with various $\dot{M}$'s.
We first create a steady burning model for each mass by setting an $\dot{M}$ in the middle of the steady burning zone from \cite{Piersanti2014a}, calculating through the first burst and allowing the burning rate to stabilize.
The example case shown in Figures \ref{fig:6} and \ref{fig:7} is for a $1.25 M_\odot$ model that starts with $\dot{M}_{\rm WD}=3.5\times10^{-6} M_\odot\text{ yr}^{-1}$.
Then we lower the accretion rate until the burning rate, $L_{\rm nuc}$, begins to oscillate by more than 15 per cent, and use this accretion rate as the lower steady burning boundary.

The lower stability boundary exists because at lower accretion rates, the helium shell has a lower temperature and higher density. 
When the accretion rate is lowered beneath the lower stability boundary, the heating timescale drops below the time required for the shell to adjust its thermal structure and a temporary runaway occurs. 
Above this boundary, the shell is hotter and less dense such that the thermal structure can adjust quickly enough in response to helium ignition, so nuclear burning rates are able to stabilize, as shown in Figure \ref{fig:6}.

To find the upper steady burning boundary, we raise the accretion rate until the surface radius experiences a rapid expansion.
Increasing the accretion rate discontiously, as we have done here, will naturally increase the surface radius, but only by a factor of order unity.
Due to the core mass-luminosity relation \citep{Paczynski1971}, there exists a maximum luminosity at which it can burn, and therefore a maximum accretion rate \citep{Shen2007}.
Once the accretion rate increases above this threshold, mass builds up in the shell and the WD either enters the RG regime \citep{Piersanti2014a} or drives an optically thick wind \citep{Hachisu1999}, both resulting in the rapid expansion of the surface radius by a few orders of magnitude, as shown in Figure \ref{fig:7}.

\subsection{He Burning during Binary Evolution}

\begin{figure}[H]
  \centering
  \includegraphics[width = \columnwidth]{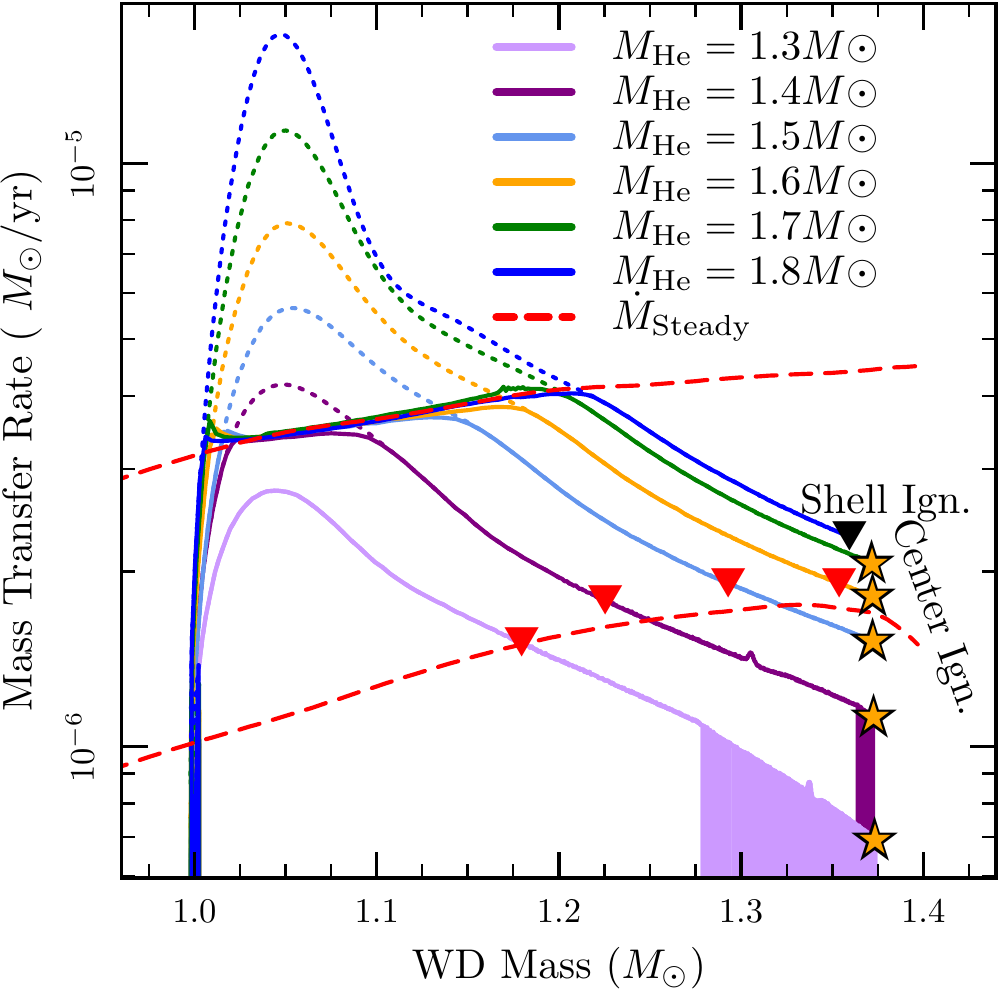}
  \caption{\footnotesize Evolutionary tracks of the mass transfer rates for six systems with different initial donor masses.
  All systems start with the same WD accretor model ($M_{\rm{WD}}=1.0 M_\odot$) and the same initial orbital period ($P_{\rm{orb},0}=3$ hours).
  The solid tracks are the rate at which the WD is gaining mass; the dotted tracks are the rate at which the He star is losing mass.
  The difference between the dotted and solid tracks represent the mass that is lost from the system. 
  The red dashed lines border the steady helium burning regime, as found in \S \ref{sec:bound}.
  The downward-facing red triangles mark where the helium burning rate begins to oscillate by more than a factor of two.}
  \label{fig:1}
\end{figure}

While the system is in the regime for steady He-burning, the WD is burning helium to carbon and oxygen at the same rate that it is accreting helium. 
As the mass transfer rates rise above the steady burning regime, the WD rapidly expands into its Roche lobe and only accepts mass at the maximum steady burning rate.
The rest of the mass is lost from the system such that $\dot{M}_{\rm WD} + \dot{M}_{\rm wind} = -\dot{M}_{\rm He}$, where $\dot{M}_{\rm wind}$ is the rate of mass loss from the binary system, as shown by the difference between the dotted and solid lines in Figure \ref{fig:1}.

We compute $\dot{M}_{\rm wind}$ by attenuating the mass transfer efficiency as the WD expands into a significant fraction of its Roche lobe. 
The shape of the mass transfer efficiency versus $R_{\rm WD}/R_{\rm RL}$ has little effect on the $\dot{M}$ accepted by the WD due to the high sensitivity of the WD radius to the accretion rate near the upper stability boundary, as shown in Figure \ref{fig:7}. 
The dynamic range of radii within the steady helium burning range is relatively small compared to that near the upper steady boundary. 
We compare this method with that used in \cite{Yoon2003} who used a radiation driven wind using $\dot{M}=10^{-2}R_{\rm WD}L_{\rm WD}/[GM_{\rm WD}(1-\Gamma)]$, where $R_{\rm WD}$, $L_{\rm WD}$, and $M_{\rm WD}$ are the radius, luminosity, and mass of the WD, and $\Gamma$ is the ratio of photospheric luminosity to the Eddington luminosity. 
For their model starting with $M_{\rm WD}=1.0 M_\odot$, $M_{\rm He}=1.6 M_\odot$, when $M_{\rm WD}=1.04 M_\odot$ the wind reaches a maximum of $4.8\times10^{-6} M_\odot\text{ yr}^{-1}$.
In our model with the same initial conditions, at the corresponding point in evolution, we calculate a maximum wind rate of $4.2\times10^{-6} M_\odot\text{ yr}^{-1}$, which is comparable.

The system mass-loss increases the orbital separation, and thus acts to reduce the rapid mass transfer rates.
We assume that the mass loss from the system takes with it the specific angular momentum of the WD, and address the validity of this assumption in \S \ref{sec:angmom}.
When mass transfer rates fall back into, and then below the steady He-burning regime, the envelope begins to oscillate (in temperature, density, burning rate, etc.).
The mass transfer rates as a function of the WD accretor mass are shown in Figure \ref{fig:1} for six models with different initial donor masses.
The downward-facing red triangles mark where the helium burning rate begins to oscillate by more than a factor of two. 
They differ from the lower stability boundary due to their accretion history building a layer of hot C/O on top of the colder core.

Steady burning on the WD avoids the complications of flashes and whether the WD mass increases or decreases after multiple novae.
On the other hand it introduces the complication of a possible carbon shell ignition before a core ignition.
These high mass transfer rates and strong helium burning rapidly heat the outer-most carbon in the WD.
So much so that if mass transfer rates stay too high for too long, the WD experiences a carbon shell ignition that \citep{Nomoto1985, Piersanti2014a} propagates inwards and transforms the C/O WD into a O/Ne WD.
The situation that results is a ``race'' between the core and the shell as to which will ignite first.

Both the WD and helium star models were first constructed in single star evolution, with the only simulated binary interaction being the artificial removal of their envelopes at the last stage in their evolution before becoming a WD or helium star.
After their envelopes have been removed, and the WD cools for 10 Myr, they are placed into a 3 hour orbit in a binary run.
All simulation after this point are true binary runs whose angular momentum evolution is only affected by gravitational wave radiation (GWR) and mass loss as described in \S \ref{sec:angmom}.

The mass transfer rate from the helium star due to Roche lobe overflow (RLOF) is computed using the ``Ritter'' implicit scheme of $\texttt{MESA}$ \citep{Paxton2015a}, which computes the prescription given by \cite{Ritter1988}.

\subsection{Angular Momentum Loss in Winds}\label{sec:angmom}

We now address the validity of the assumption that the mass loss from the system takes with it the specific angular momentum of the WD.
It is perfectly valid if the wind leaves the system quickly without interacting with the binary after being launched off the WD's surface.

If, however, the wind speed is low, then it gets gravitationally torqued by the stars as it leaves the system, extracting extra angular momentum from the orbits.
\cite{Hachisu1999} explore this issue of the angular momentum evolution in a binary, and parameterize the extraction of extra angular momentum from a slow wind.
They first define the dimensionless quantity $l_{\rm wind}$ as

\begin{equation}\label{eqn:8} \left(\dfrac{\dot{J}}{\dot{M}}\right)_{\rm wind} = l_{\rm wind}a^2\Omega_{\rm orb} , \end{equation}
where $J$ is the total angular momentum, $\dot{M}$ is the mass loss rate of the system, $a$ is the binary separation, and $\Omega_{\rm orb}$ is the orbital angular frequency.
If, as per our assumption, the wind from the system takes with it the specific angular momentum of the WD, then $l_{\rm wind}=(q/(1+q))^2$, where $q$ is the mass ratio $M_{\rm He}/M_{\rm WD}$.
If the wind is slow, on the other hand, it extracts more angular momentum, and thus $l_{\rm wind}$ is larger.
The expression for $l_{\rm wind}$ as given by Hachisu et al. (1999) is then

\begin{equation}\label{eqn:9} l_{\rm wind}\approx\text{max}\left[1.7-0.55\left(\dfrac{v}{a\Omega_{\rm orb}}\right)^2,\left(\dfrac{q}{1+q}\right)^2\right] , \end{equation}
where $v$ is radial velocity of the wind near the RL surface.

\begin{figure}[H]
  \centering
  \includegraphics[width = \columnwidth]{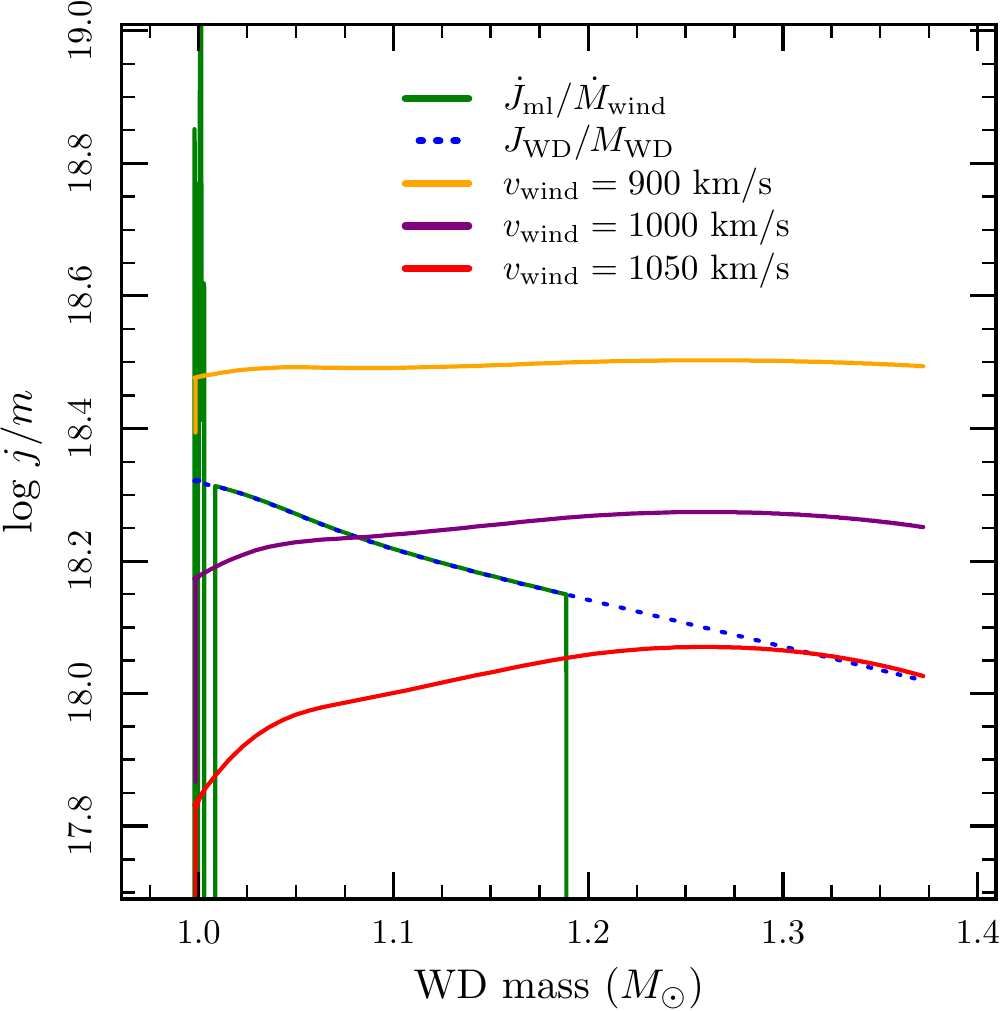}
  \caption{\footnotesize Angular momentum evolution for the binary case with $M_{\rm WD}=1.0 M_\odot$, $M_{\rm He}=1.6 M_\odot$.
  The specific angular momentum of the WD is shown by the blue dotted line. 
  The specific angular momentum of the wind leaving the system is shown by the solid green line.
  The green line drops to zero when the WD mass reaches ${\approx}1.19 M_\odot$ because the system mass loss rate drops to zero.
  The orange, purple, and red solid lines show what the specific angular momentum of the wind would be if the speeds were 900, 1000, and 1050 km/s, respectively.
  This shows that our angular momentum loss rate assumptions are valid if $v_{\rm wind}\gtrsim1000$km/s.}
  \label{fig:16}
\end{figure}

We show in Figure \ref{fig:16} for the case with $M_{\rm WD}=1.0 M_\odot$, $M_{\rm He}=1.6 M_\odot$, the specific angular momentum of the WD with the blue dotted line, and the specific angular momentum of the wind leaving the WD surface with solid green. 
The orange, purple, and red solid lines are from equation \ref{eqn:8} assuming the left-hand-side option in the square brackets in equation \ref{eqn:9} with wind speeds of 900, 1000, and 1050 km/s, respectively. 
This figure shows that if $v_{\rm wind}\gtrsim1000$ km/s, then our initial assumption that the mass loss from the system takes with it the specific angular momentum of the WD is valid. 
Lower wind velocities would extract extra angular momentum, thus decreasing the orbital period and increasing system mass loss rates.
We did not consider that possibility for this initial exploration.

\section{Core and Envelope Evolution}\label{sec:core}

Here we return to describe the ``race'' between the core and the envelope as to which will ignite carbon first.
We present example models that explore the different possible outcomes, and thus the final fate of these systems.

\begin{figure}[H]
  \centering
  \includegraphics[width = \columnwidth]{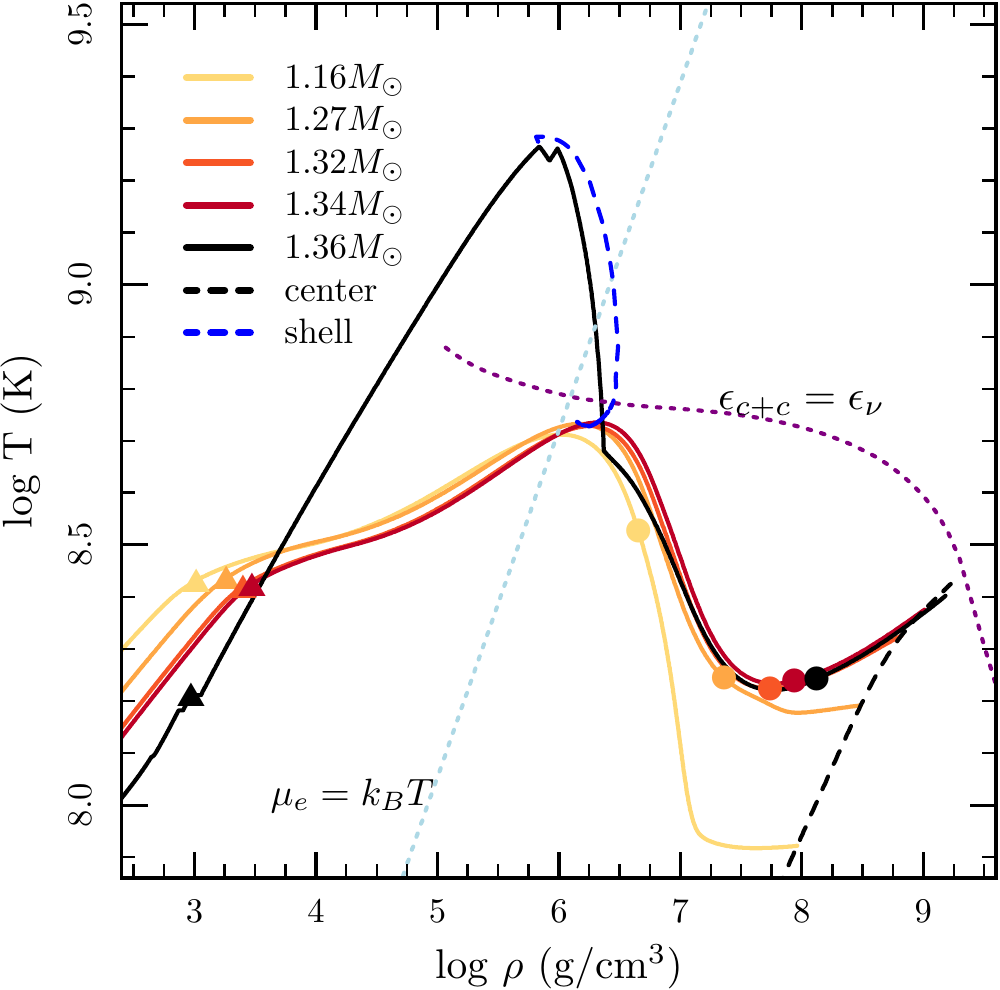}
  \caption{\footnotesize Binary system with $M_{\rm He}=1.8 M_\odot$. 
  Solid lines show the profiles of the WD, labeled by the mass at that point in their evolution.
  They move in $\rho-T$ space towards the $\epsilon_{C+C}=\epsilon_\nu$ curve, but due to the high accretion rate the envelope crosses the curve first and the WD experiences a carbon shell ignition at the mass coordinate of $1.349 M_\odot$, nearly at the surface. 
  The triangles mark the helium burning zone, and the circles mark the mass coordinate $m_r=1.0 M_\odot$.}
  \label{fig:2}
\end{figure}

\begin{figure}[H]
  \centering
  \includegraphics[width = \columnwidth]{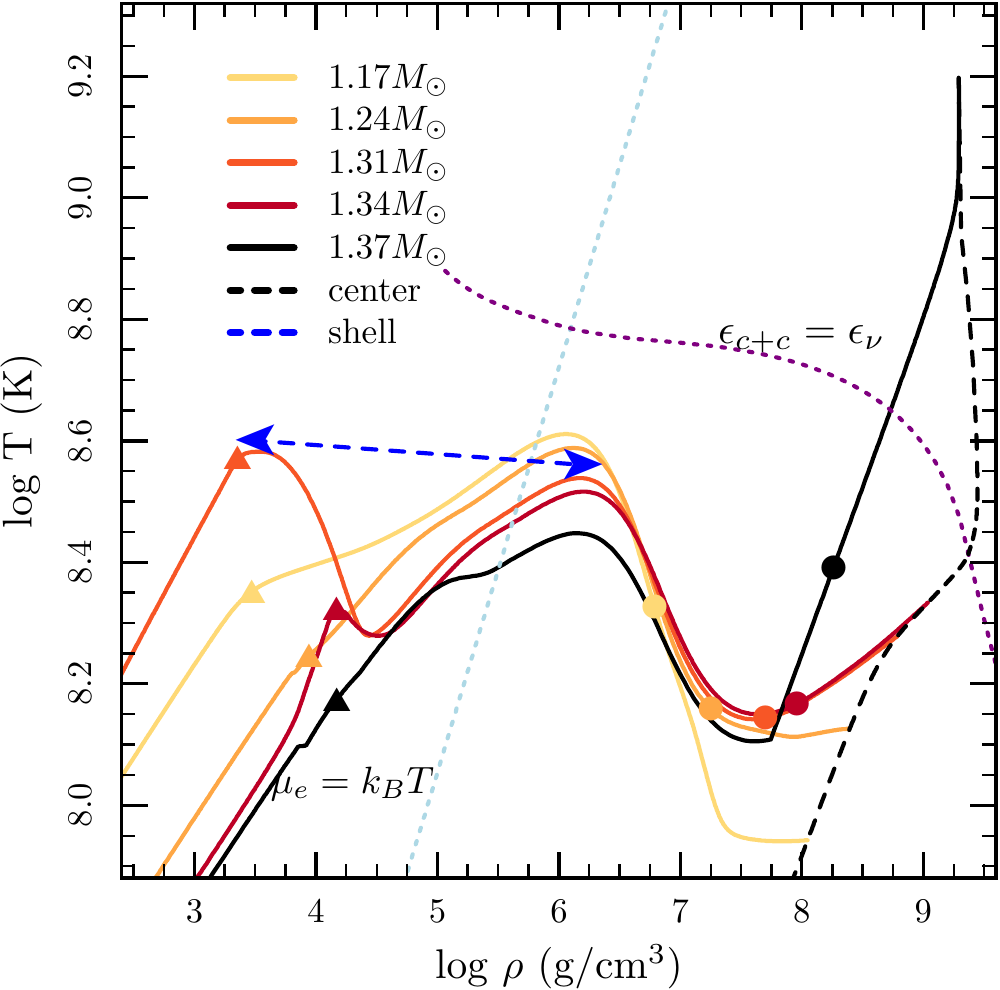}
  \caption{\footnotesize Profiles of the WD accretor model with $1.3 M_\odot$ helium star donor. 
  The triangles mark the helium burning zone, and the circles mark the mass coordinate $m_r=1.0 M_\odot$. 
  The last few profiles show the envelope oscillating due to mass transfer rates falling below the steady He-burning regime.}
  \label{fig:3}
\end{figure}

We start with the case with the highest mass donor, and thus the highest accretion rates for the entire accretion lifetime.
In Figure \ref{fig:2} we show the evolution of the core and the carbon shell (tracked by maximum temperature) of the model with initial donor mass $M_{\rm{He}}=1.8 M_\odot$.
This case shows that high accretion rates dump more heat into the envelope than the compressionial heat in the core, and the shell crosses $\epsilon_{C+C}=\epsilon_\nu$ before the core.
Shell carbon ignition occurs when the WD reaches a mass of $1.360 M_\odot$ at a mass coordinate of $1.349 M_\odot$, nearly at the surface.

On the low mass donor end, however, carbon may not ignite at all.
The model with initial donor mass $M_{\rm{He}}=1.3 M_\odot$ stays at lower accretion rates and falls far enough below the steady helum burning range that the mild helium oscillations become powerful enough to blow off mass. 
The blue dashed line shows how the maximum temperature in the shell will oscillate between the flashing helium burning shell and the hot, compressed C/O ashes below the burning layer.
At these accretion rates, however, the accretion efficiency stays above 70\%, allowing the WD to gain enough mass to trigger carbon ignition in the core. 
We compare our accreted mass retention efficiency calculated in the final stages of this model using super-Eddington winds to the analytical

\begin{figure}[H]
  \centering
  \includegraphics[width = \columnwidth]{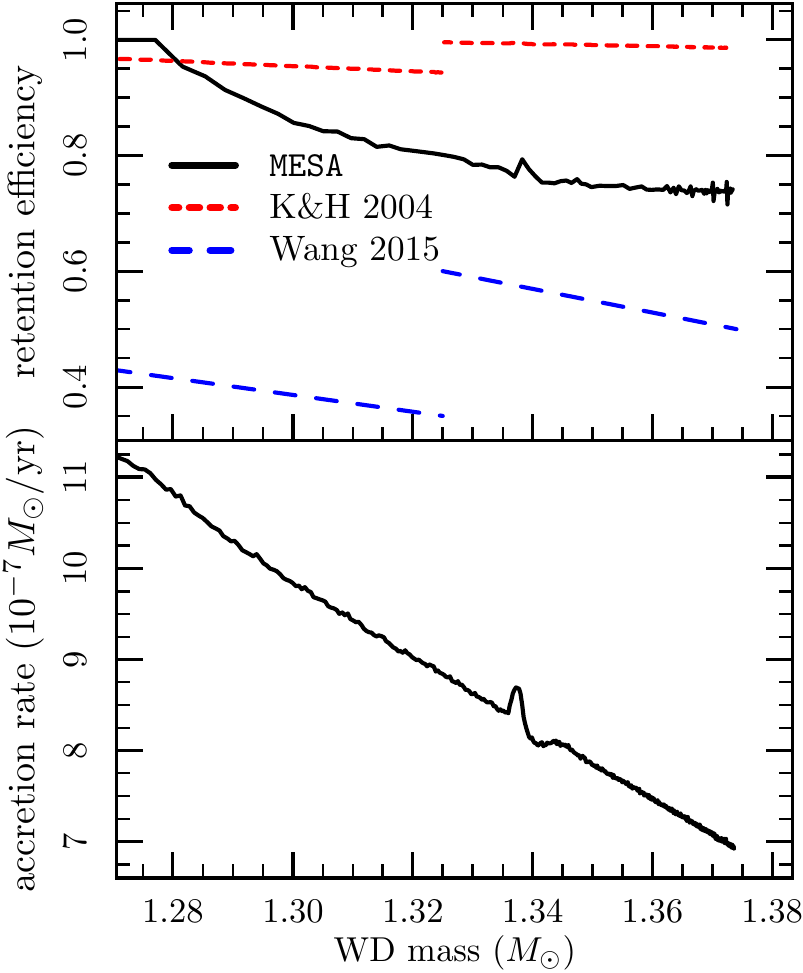}
  \caption{\footnotesize The top panel shows the accreted mass retention efficiency during the last stages of the system with the $1.3 M_\odot$ donor stars.
  The black line shows what we calculated by taking the ratio of the growth rate of the accretor and the mass loss rate of the donor.
  The red short-dashed lines show the retention efficiency given analytically by \cite{Kato2004} for 1.3 and $1.35 M_\odot$ cores.
  The blue long-dashed lines show the retention efficiency given analytically by \cite{Wang2015} for 1.3 and $1.35 M_\odot$ cores.
  The bottom panel shows the mass transfer rate being donated to the WD.}
  \label{fig:4}
\end{figure}

\noindent estimates of \cite{Kato2004} and \cite{Wang2015} in Figure \ref{fig:4}.
\cite{Wang2015} used $\texttt{MESA}$ (r3661) for their study, but used the default super-Eddington wind settings, whereas we triggered super-Eddington winds while the WD is still in a compact configuration in order to speed the computation.
Systems with donor mass $M_{\rm{He}}\lesssim1.3 M_\odot$, therefore, will fall to lower accretion rates and efficiencies, and ultimately lose contact before a core ignition occurs. 
The components, now both C/O WDs, will spiral in due to GWR and contribute to the double-degenerate channel for SNe Ia, as discussed in \S \ref{sec:lowmass}. 

\begin{figure}[H]
  \centering
  \includegraphics[width = \columnwidth]{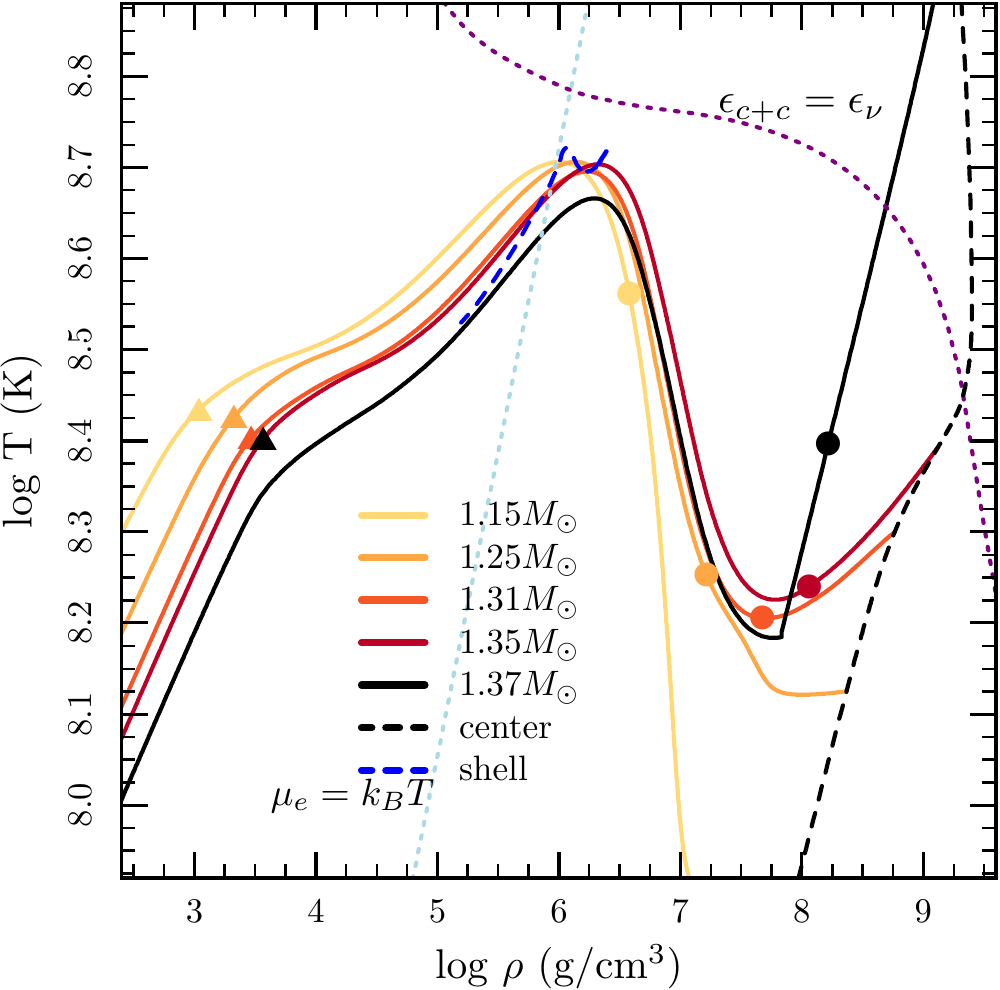}
  \caption{\footnotesize Profiles of $\rho-T$ of the WD in the binary system with $M_{\rm He}=1.6 M_\odot$.
  Markings have the same meaning as in Figure \ref{fig:3}.}
  \label{fig:8}
\end{figure}

\begin{figure}[H]
  \centering
  \includegraphics[width = \columnwidth]{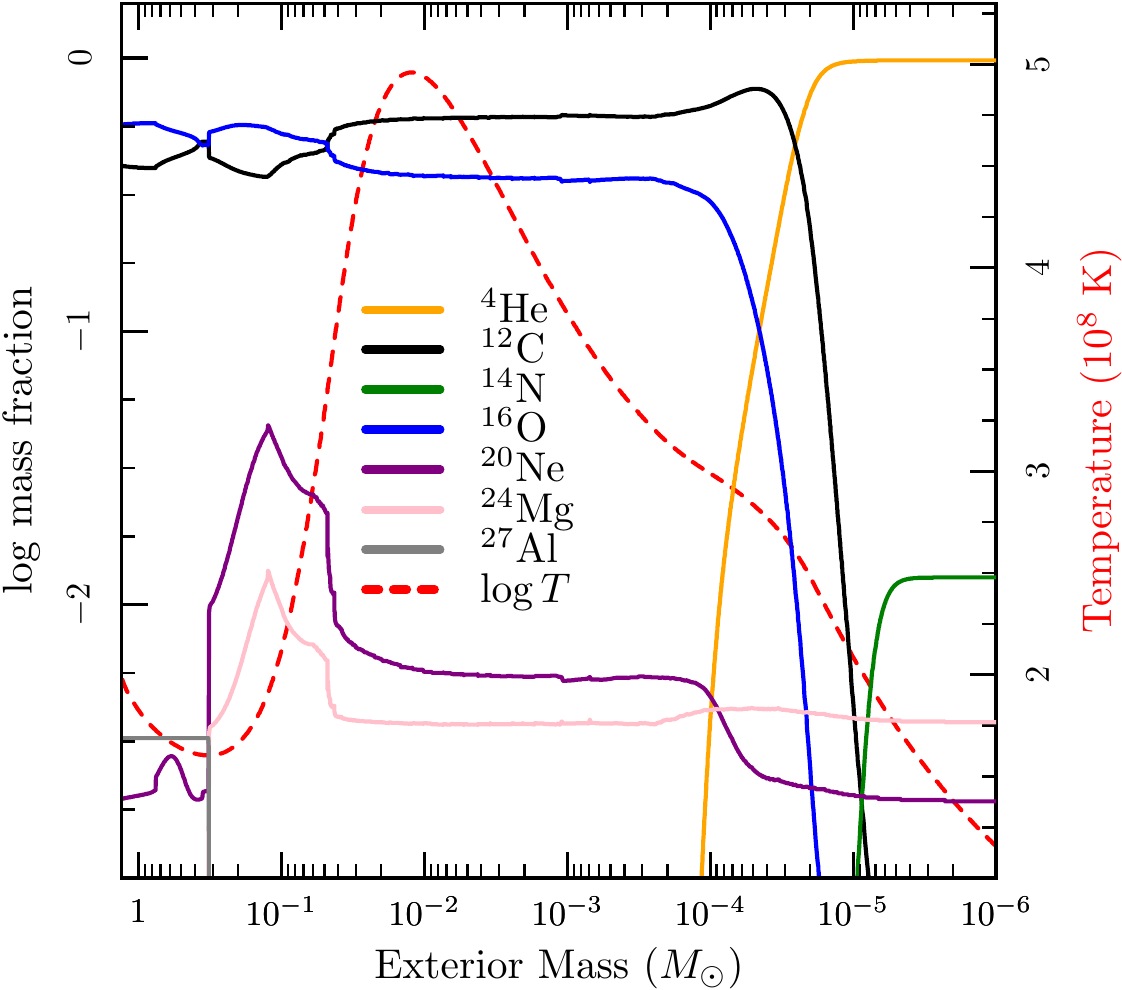}
  \caption{\footnotesize Profile of abundance on the left y-axis and temperature in units of $10^8$ K on the right y-axis.
  This is from the WD in the binary system with $M_{\rm He}=1.6 M_\odot$, using the same profile as the middle profile in Figure \ref{fig:8}, when $M_{\rm WD}=1.31 M_\odot$.}
  \label{fig:9}
\end{figure}

Between the high and low mass donor limits is the region where core carbon ignitions occur.
This is where the donor mass is high enough that the accretion rates do not fall too far below the stability boundary before $M_{\rm Ch}$ is reached, but also low enough that it avoids a carbon shell ignition from long sustained high accretion rates.
A clear example of this is shown in Figure \ref{fig:8}, with the trajectories of the core and envelope shown with profiles in $\rho-T$ space for a system with $M_{\rm{He}}=1.6 M_\odot$.
In Figure \ref{fig:9} we show the composition and temperature profile when this model has grown to $1.31 M_\odot$, the same profile as that in Figure \ref{fig:8}.
As the C/O envelope is $0.31 M_\odot$, this shows that the base is at a temperature minimum, and the temperature maximum in the envelope is only ${\approx}10^{-2} M_\odot$ deep.
The core ignition in these models indicates that they are likely SN Ia progenitors.

\section{Deviations From Adiabatic Core Compression}\label{sec:dev}

The black dashed lines in the previous figures (Figures \ref{fig:2}, \ref{fig:3}, \ref{fig:8}) show the WD core trajectory, which, at $T_c\approx 2\times 10^8$K, deviates from the adiabatic compression expected for such high $\dot M$'s. 
Since there is no time for heat transfer at these rapid accretion rates, it is the onset of neutrino cooling  \citep{Paczynski1971} at a rate $\epsilon_\nu$ that leads to this deviation. 
This impacts the shell-core race to carbon ignition and also hints at a possible sensitivity to the initial WD core temperature, which depends on the WD age at the time accretion starts.

\begin{figure}[H]
  \centering
  \includegraphics[width = \columnwidth]{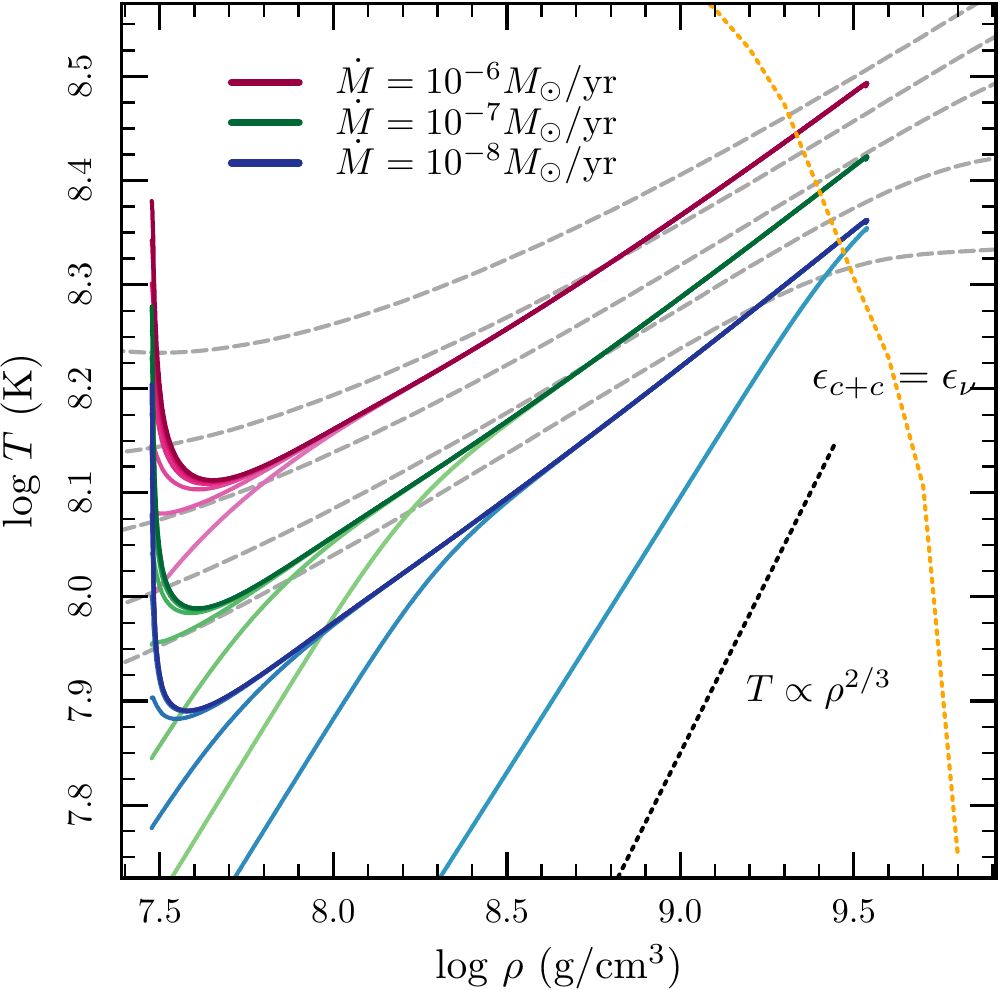}
  \caption{\footnotesize Evolution of core of WD given different $\dot M$'s and starting core temperatures.
	The grey dashed lines are lines of constant neutrino cooling timescales from $3\times10^4$ to $3\times10^6$ years, spaced logarithmically, from top to bottom. 
	As always, the dotted line represents $\epsilon_{\rm{c+c}}=\epsilon_\nu$ for $\{X_{C12}=0.5,X_{O16}=0.5\}$.}
  \label{fig:5}
\end{figure}

It is when the the neutrino cooling timescale, $t_\nu=c_pT/\epsilon_\nu$, approaches the compressional timescale $t_{\rm comp}=(d \ln \rho/dt)^{-1}$ that the core no longer evolves adiabatically. 
To derive the resulting relations more rigorously, we start by writing the entropy change, $ds$, in the form of equation A6 of \cite{Townsley2004},
\begin{equation}
\label{eqn:1} ds = \dfrac{k_B}{\mu m_p}(a\,d\ln T - b\,d\ln\rho),
\end{equation}
where $k_B$ is Boltzmann's constant, $\mu$ is the mean molecular weight, $m_p$ is the proton mass, $a=1.22+0.41\Gamma^{1/3}$, $b=0.91+0.14\Gamma^{1/3}$, and $\Gamma = (Ze)^2/akT$ where $a$ is the ion separation, and we set $Z=6.857$ for an equal (by mass) mix of carbon and oxygen.
We divide each side by $dt$ and multiply by $T$ to obtain
\begin{equation}
\label{eqn:2} -\epsilon_\nu(\rho,T) = T\dfrac{ds}{dt} = \dfrac{k_BT}{\mu m_p}\left(a\dfrac{d\ln T}{dt} - b\dfrac{d\ln\rho}{dt}\right),
\end{equation}
and then expand the second term on the right as
\begin{equation}
\label{eqn:3} \dfrac{d\ln\rho}{dt} = \dfrac{d\ln\rho}{d\ln M}\dfrac{d\ln M}{dt}.
\end{equation}
We use $\texttt{MESA}$ to compute the steeply rising value of $n=d\ln\rho/d\ln M$ as $M\to M_{\rm{Ch}}$. 
For a constant $\dot{M}$, we can then solve the resulting differential equation for $T(t)$ given an initial temperature.

This results in evolutionary trajectories in the $\rho-T$ plane that are shown in Figure \ref{fig:5} to converge to a common $\rho-T$ line when neutrino cooling is dominant. 
This common line is one where the neutrino cooling timescale equals that of compression (shown as light grey curves in Figure \ref{fig:5}).
Cold initial conditions adiabatically rise in temperature until meeting the $\dot{M}$-dependent trajectory, whereas hot initial conditions will cool to reach the attractor. 
Hence, a range of initial conditions will converge to the same trajectory given the same $\dot{M}$. 
The same convergence of the evolution of the central temperature and density was shown in \cite{Paczynski1971} in the context of intermediate mass AGB stars.
\cite{Paczynski1971}, however, only show one convergence line due to the core mass-luminosity relation.
Much colder cores with lower accretion rates never reach regimes where neutrino cooling is strong enough to cause convergence, and so behave differently.

\section{HR Diagrams}\label{sec:hr}

\begin{figure}[H]
  \centering
  \includegraphics[width = \columnwidth]{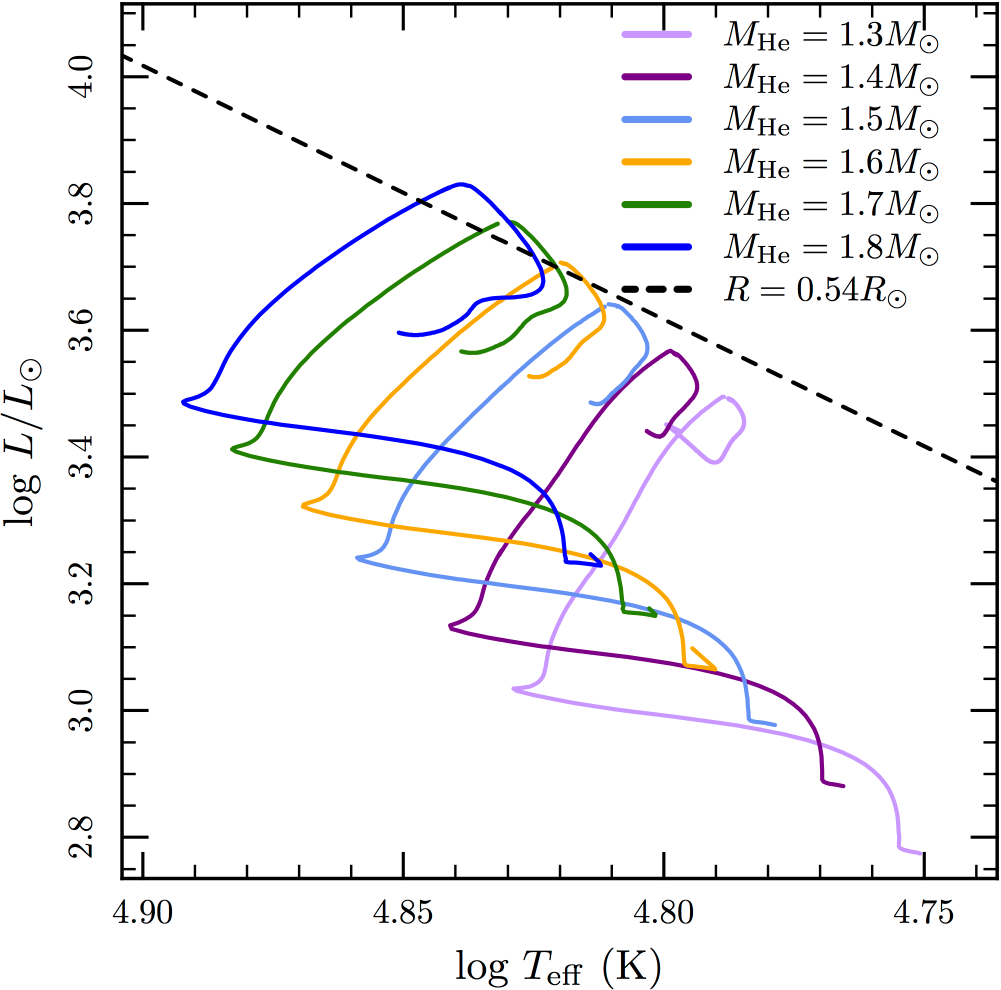}
  \caption{\footnotesize HR diagram of the donors of masses $1.3-1.8 M_\odot$. 
  Evolution starts with core helium ignition in the lower, right-hand corner, evolves through to core helium exhaustion at the left-most (hottest) point, then expands and brightens with shell helium burning, and reaches the brightest point when mass transfer starts. 
  A line of constant radius near the initial Roche lobe radius for these systems, $R=0.54 R_\odot$, is shown by the black dashed line.}
  \label{fig:17}
\end{figure}

We now explore the observability of these systems by following their trajectories through the HR diagram. 
We show the evolution of the donors in Figure \ref{fig:17}, starting from ignition of helium in the core in the lower right-hand corner, and evolving to higher temperatures and luminosities along the helium MS. 
Once helium in the core has been exhausted, shown by the left-most (hottest) point in the evolution, the envelopes of the helium donors begin to expand due to helium shell burning, causing $T_{\rm eff}$ to decrease and the luminosity to increase. 
The rise in 

\noindent luminosity is halted by the start of mass transfer as the helium stars' envelopes expand into their RLs. 
The position of the donors in the HR diagram when the carbon ignites falls right in the most heavily populated region of Figure 4 of \cite{Wang2014}, which plots distribution of donor stars in systems that achieved a core ignition through the helium star channel. 

With initial orbital periods ($P_{\rm orb,0}$) that are longer (shorter), the donors will have more (less) time and room to expand their envelopes until RLOF, leading to donor stars that are redder (bluer) and brighter (dimmer) when carbon ignites in the WD.
This suggests a diagonal patch in the HR diagram for the state of donors.
\cite{Wang2009} and \cite{Liu2010} find systems in which the donor star donated enough helium that the helium shell burning layer starts becoming exposed, leading to a rapid evolution to higher temperatures and luminosites. 
This requires systems with either lower mass donors, which will fall below the steady helium burning mass transfer rates, or lower mass WDs, which means that the WD will spend a longer time accreting and thus has a much higher likelihood of experiencing a carbon shell ignition before a core ignition. 

\begin{figure}[H]
  \centering
  \includegraphics[width = \columnwidth]{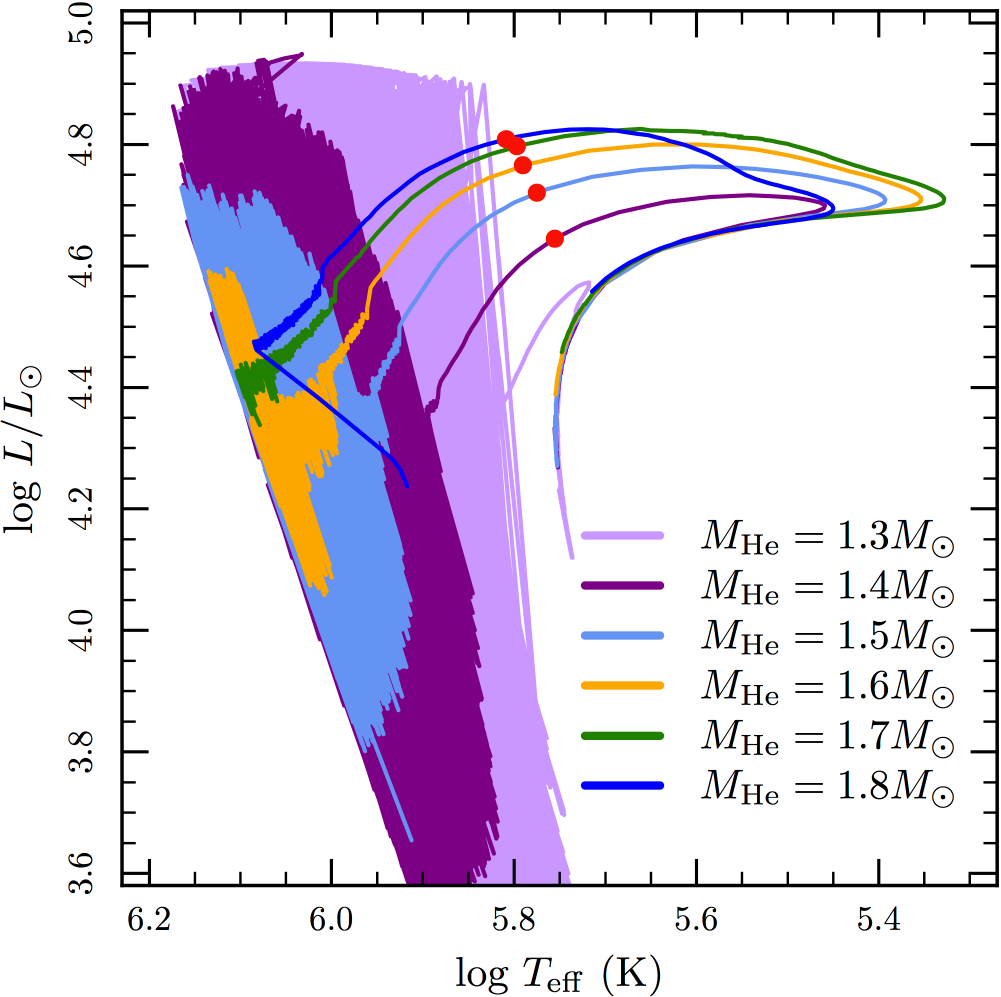}
  \caption{\footnotesize HR diagram of the WD accretors given six different donor masses, corresponding to the same colored lines in Figure \ref{fig:17}. 
  Evolution started in the middle of the plot, moves right (hotter) then left (cooler). 
  The red markers denote where the optically thick wind phase ends. 
  The WDs with the $1.5-1.6 M_\odot$ helium star donors enter into mild helium oscillations (no mass loss) at the end of their evolution, while the WDs with the $1.3-1.4 M_\odot$ helium star donors will experience oscillations strong enough to remove mass.}
  \label{fig:18}
\end{figure}

We also show, in Figure \ref{fig:18}, the evolution of the WD accretors in the HR diagram. 
Due to the steady helium burning on the surface of the WDs, they are hot and bright enough to be classified as supersoft X-ray sources (SSS) \citep{Kahabka1997, Iben1994}. 
The WDs begin their evolution in the middle of Figure \ref{fig:18} at about $\log T_{\rm eff}=5.75$ and $\log L/L_\odot=4.4$, but will be obscured by an optically thick wind until they reach the red marker, where mass transfer rates have decreased below the upper steady helium burning boundary and the transfer efficiency reaches unity. 
For the systems with the lowest mass donors ($M_{\rm He}=1.3-1.5 M_\odot$), the mass transfer rates drop below the steady helium burning range near the end of their evolution. 
The range of oscillations of $T_{\rm eff}$ during the mild helium oscillations is less than a factor of two. 
The luminosity, however, changes by about an order of magnitude, on the timescale of ${\approx}5$ years, so this should be visible to observers.

Note that in all six models, by the time the carbon ignites, the optically thick wind has been inactive for tens of thousands of years, meaning that there should be no nearby circumstellar material.

\section{Lower Mass Binary Case Leading to WD Mergers}\label{sec:lowmass}

\begin{figure}[H]
  \centering
  \includegraphics[width = \columnwidth]{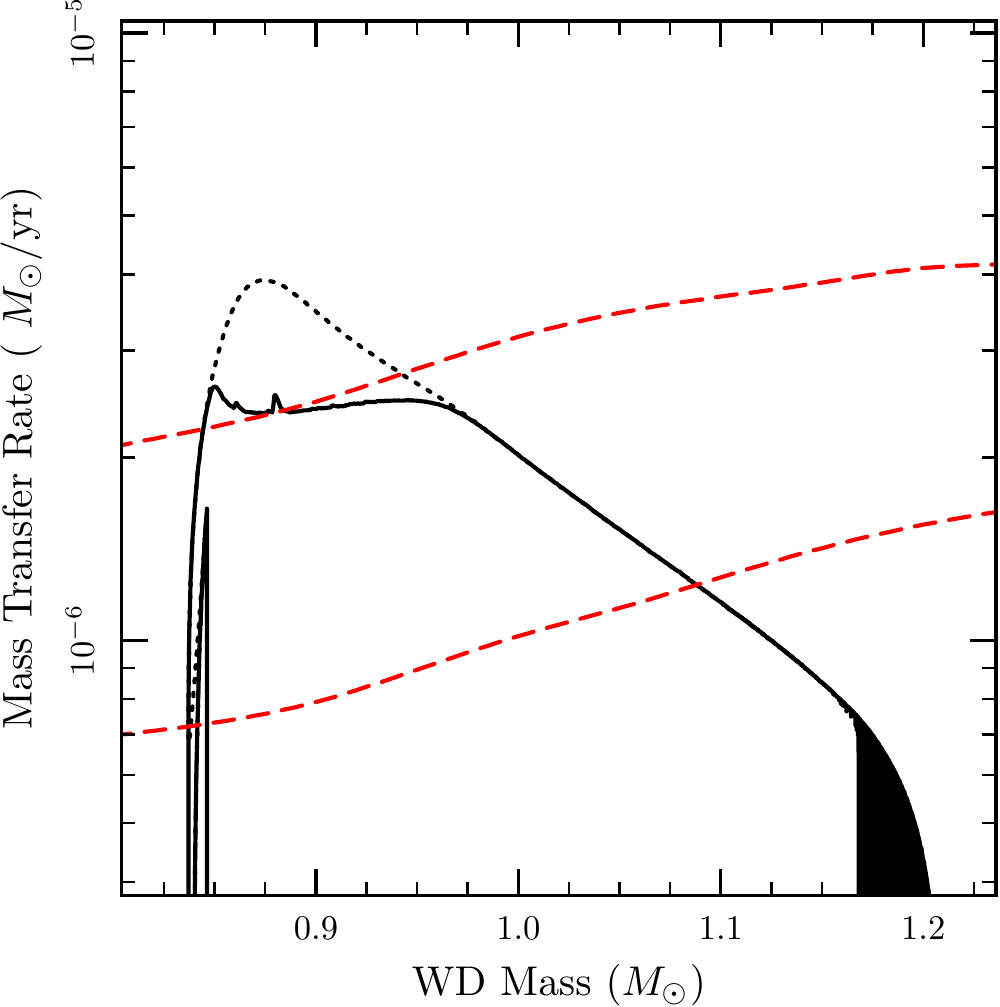}
  \caption{\footnotesize The mass transfer history for the low mass case. 
  The WD accretes within the steady helium burning range until $M_{\rm WD}>1.09 M_\odot$, but only stops conserving mass when $M_{\rm WD}>1.16 M_\odot$, and grows to $M_{\rm WD}=1.21 M_\odot$.}
  \label{fig:10}
\end{figure}

We now test our prescriptions against the calculations done by \cite{Ruiter2012} in their example case in section 2.3.
They start with $5.65$ and $4.32 M_\odot$ MS stars separated by 37 $R_\odot$.
After the primary evolves to become a C/O WD and engages in a common envelope with the secondary as is exhausts core hydrogen, they are left with a $0.84 M_\odot$ WD and $1.25 M_\odot$ helium star separated by $1.74 R_\odot$.
These are the parameters we initialize, with the mass transfer history shown in Figure \ref{fig:10}. 
We model the system until the He star loses contact and the stars spiral inward due to GWR and begin to merge.
When the mass of the helium layer on the donor decreases down to $7\times10^{-3} M_\odot$, the helium burning layer becomes too weak to support an extended convective envelope and the surface contracts below the RL.
Compared to \cite{Ruiter2012}, we calculate less system mass loss, resulting in slightly more massive components, $M_{\rm WD}=1.21 M_\odot$, $M_{\rm He}=0.81 M_\odot$, compared to $M_{\rm WD}=1.19 M_\odot$, $M_{\rm He}=0.77 M_\odot$ for \cite{Ruiter2012}.
This also leads to a shorter merger time, 625 Myr for our work and 1130 Myr for \cite{Ruiter2012}, after loss of contact.

As \cite{Ruiter2012} used the accumulation efficiencies of \cite{Kato2004} for helium accretion onto WDs, this serves as an additional comparison (along with Figure \ref{fig:4}) to the mass transfer prescriptions of \cite{Kato2004}.
This shows that for WDs in the $1.2 M_\odot$ range the \cite{Kato2004} prescription match our ending masses to with a few per cent, but somewhat larger discrepancies appear for masses closer to $M_{\rm Ch}$ (Figure \ref{fig:4}).

\section{Conclusions}\label{sec:conc}

We ran binary simulations using $\texttt{MESA}$ for WD + Helium star systems, calculating stellar structure equations for both stars simultaneously, along with the binary parameters that take into account their interaction through mass transfer. 
The systems we study include $1.0 M_\odot$ C/O WDs in 3 hour orbital periods ($P_{\rm orb}$) with helium stars with masses $1.3 M_\odot\le M_{\rm{He}} \le 1.8 M_\odot$. 
This configuration allows the steady growth of the WD core mass via steady helium burning on the WD surface. 
At mass transfer rates above the steady helium burning range, we assume an optically thick wind that ejects all the donated mass above the maximum steady helium burning rate from the system. 
Below the steady helium burning range the WD begins mild helium oscillations. 

During the initial turn-on of mass transfer, the WD surface is initially cold and must be heated through several helium flashes until steady helium burning can begin. 
Using Figure 10 of \cite{Shen2014a}, we conclude that all the first bursts on the $1.0 M_\odot$ WD models are well below the detonation threshold, but the first burst on the lower mass $0.84 M_\odot$ WD may support a detonation. 

We assume that the mass that is lost from the system takes with it the specific angular momentum of the accretor, which is justified as long as the wind speeds exceed $v_{\rm wind}\gtrsim 1000$km/s. 
We note here that some orbital angular momentum may be lost in spinning up the accreting WD, which we did not explore in this study. 
During the optically thick wind phase, however, the accretion stream coming through L1 may not be able to form a disk in the extended RG envelope around the WD, and may become part of that envelope once it reached a depth at which the surrounding density matches the stream density, thus preventing any angular momentum transfer to the WD. 
This way, the angular momentum of the accretion stream gets fed back into the orbit through tidal effects on the extended envelope. 
We leave this subject, as well as internal angular momentum transport, to future studies.

The systems with the lowest mass donors ($M_{\rm He}=1.3-1.4 M_\odot$) begin mild helium burning oscillations before the WD reaches $M_{\rm Ch}$. 
Systems with $1.3 M_\odot \lesssim M_{\rm He} \lesssim 1.7 M_\odot$ experience a core ignition in a $M_{\rm Ch}$ WD.
Systems with $M_{\rm He}\gtrsim 1.8 M_\odot$ experience a shell ignition of carbon in the helium burning ashes. 
The corresponding maximum helium star donor mass for lower initial mass WDs might be even lower, given that they must spend more time accreting to grow to $M_{\rm Ch}$, but the birthrate from those systems, from Table 1 in \cite{Wang2009}, is negligible compared to systems with higher mass WDs. 
This shell ignition is non-explosive and will lead to a thin carbon burning front that will propagate through the C/O WD all the way to the center, converting a C/O WD into an O/Ne WD \citep{Nomoto1985, Saio1985, Timmes1994, Saio1998}. 
Without any carbon in the center, when the WD mass nears $M_{\rm Ch}$, electron captures onto $^{24}$Mg and $^{20}$Ne will remove pressure support from the core, resulting in an accretion induce collapse (AIC), leaving behind a neutron star \citep{Miyaji1980, Schwab2015a}. 
In future studies, we intend to run simulations over a large range of initial orbital periods and WD masses to determine the real upper boundary for the He star donor mass that allows a core ignition in the $\log P_{\rm orb,0}-M_{\rm He}$ plane in Figure 7 in \cite{Wang2009}.
In addition to changing the expected SN Ia rate, this would yield an AIC rate for this new channel (Brooks et al. 2016 in preparation). 
This would be in addition to the rate expected from the initially more massive, presumably O/Ne WD accretors.

Systems with $M_{\rm WD}\gtrsim1.1 M_\odot$, although assigned a much higher contribution to the SNe Ia birthrate than their lower mass WD counterparts, are not likely to be C/O WDs \citep{Nomoto1984, Timmes1994, Ritossa1996, Garcia-Berro1997}. 
\cite{Wang2014} explore systems with so-called ``hydrid'' C/O/Ne WDs that experience an off-center carbon ignition during WD formation that was quenched by convective boundary mixing and left an unburnt carbon oxygen core of up to $0.3 M_\odot$, surrounded by an O/Ne mantle of equal or greater mass \citep{Denissenkov2013, Chen2014, Denissenkov2015} (This opens the possibility that carbon shell ignitions on C/O WDs may experience flame quenching before converting the entire core into O/Ne). 
This thick O/Ne mantle would prevent carbon flames that ignited in the ashes from steady helium burning from reaching the unburnt core carbon. 
Therefore, carbon shell ignitions, like the one experienced by the WD with the $M_{\rm He}=1.8 M_\odot$ in this study, will not prevent hybrid C/O/Ne WDs from achieving core carbon ignition. 
It is unlikely, however, that this sort of WD can lead to a normal SN Ia because a deflagration ignited in the C/O core cannot transition into a detonation in the O/Ne mantle easily because the critical mass for a detonation of an O/Ne mixture is much larger than that of carbon \citep{Denissenkov2015}. 
These hybrid C/O/Ne WDs are more likely to lead to the subclass SN Iax \citep{Fink2013, Kromer2015}.

We also found that, due to rapid neutrino cooling in warm and dense cores, evolution of the central temperature and density will converge to accretion-rate dependent trajectories until ignition, given a high enough initial central temperature.

We acknowledge stimulating workshops at Sky House where these ideas germinated.
We thank Ken Shen for helpful discusions regarding binary configurations.
This work was supported by the National Science Foundation under grants PHY 11-25915, AST 11-09174, and AST 12-05574.
Josiah Schwab is supported by the National Science Foundation Graduate Research Fellowship Program under Grant No. DGE 11-06400 and by NSF Grant No. AST 12-05732.
Most of the simulations for this work were made possible by the Triton Resource, a high-performance research computing system operated by the San Diego Supercomputer Center at UC San Diego.

\bibliographystyle{apj}
\bibliography{highmass_donor}

\begin{thebibliography}{42}
\expandafter\ifx\csname natexlab\endcsname\relax\def\natexlab#1{#1}\fi

\bibitem[{Chen {et~al.}(2014)Chen, Herwig, Denissenkov, \& Paxton}]{Chen2014}
Chen, M.~C., Herwig, F., Denissenkov, P.~A., \& Paxton, B. 2014, Monthly
  Notices of the Royal Astronomical Society, 440, 1274

\bibitem[{Denissenkov {et~al.}(2013)Denissenkov, Herwig, Truran, \&
  Paxton}]{Denissenkov2013}
Denissenkov, P.~A., Herwig, F., Truran, J.~W., \& Paxton, B. 2013, The
  Astrophysical Journal, 772, 37

\bibitem[{Denissenkov {et~al.}(2015)Denissenkov, Truran, Herwig, Jones, Paxton,
  Nomoto, Suzuki, \& Toki}]{Denissenkov2015}
Denissenkov, P.~A., Truran, J.~W., Herwig, F., Jones, S., Paxton, B., Nomoto,
  K., Suzuki, T., \& Toki, H. 2015, Monthly Notices of the Royal Astronomical
  Society, 447, 2696

\bibitem[{Fink {et~al.}(2013)Fink, Kromer, Seitenzahl, Ciaraldi-Schoolmann,
  Ropke, Sim, Pakmor, Ruiter, \& Hillebrandt}]{Fink2013}
Fink, M., Kromer, M., Seitenzahl, I.~R., Ciaraldi-Schoolmann, F., Ropke, F.~K.,
  Sim, S.~A., Pakmor, R., Ruiter, A.~J., \& Hillebrandt, W. 2013, Monthly
  Notices of the Royal Astronomical Society, 438, 1762

\bibitem[{Garc{\'{\i}}a-Berro {et~al.}(1997)Garc{\'{\i}}a-Berro, Ritossa, \&
  Iben}]{Garcia-Berro1997}
Garc{\'{\i}}a-Berro, E., Ritossa, C., \& Iben, I. 1997, The Astrophysical
  Journal, 485, 765

\bibitem[{Hachisu {et~al.}(1996)Hachisu, Kato, \& Nomoto}]{Hachisu1996}
Hachisu, I., Kato, M., \& Nomoto, K. 1996, The Astrophysical Journal, 470, L97

\bibitem[{Hachisu {et~al.}(1999)Hachisu, Kato, \& Nomoto}]{Hachisu1999}
---. 1999, The Astrophysical Journal, 522, 487

\bibitem[{Iben \& Tutukov(1989)}]{Iben1989}
Iben, Icko, J. \& Tutukov, A.~V. 1989, The Astrophysical Journal, 342, 430

\bibitem[{Iben \& Tutukov(1994)}]{Iben1994}
---. 1994, The Astrophysical Journal, 431, 264

\bibitem[{Kahabka \& van~den Heuvel(1997)}]{Kahabka1997}
Kahabka, P. \& van~den Heuvel, E. P.~J. 1997, Annual Review of Astronomy and
  Astrophysics, 35, 69

\bibitem[{Kato \& Hachisu(2004)}]{Kato2004}
Kato, M. \& Hachisu, I. 2004, The Astrophysical Journal, 613, L129

\bibitem[{Kromer {et~al.}(2015)Kromer, Ohlmann, Pakmor, Ruiter, Hillebrandt,
  Marquardt, Roepke, Seitenzahl, Sim, \& Taubenberger}]{Kromer2015}
Kromer, M., Ohlmann, S.~T., Pakmor, R., Ruiter, A.~J., Hillebrandt, W.,
  Marquardt, K.~S., Roepke, F.~K., Seitenzahl, I.~R., Sim, S.~A., \&
  Taubenberger, S. 2015, eprint arXiv:1503.04292

\bibitem[{Liu {et~al.}(2010)Liu, Chen, Wang, \& Han}]{Liu2010}
Liu, W.-M., Chen, W.-C., Wang, B., \& Han, Z.~W. 2010, Astronomy {\&}
  Astrophysics, 523, A3

\bibitem[{Miyaji {et~al.}(1980)Miyaji, Sugimoto, Nomoto, \& Yokoi}]{Miyaji1980}
Miyaji, S., Sugimoto, D., Nomoto, K., \& Yokoi, K. 1980, In: International
  Cosmic Ray Conference, 2, 13

\bibitem[{Nomoto(1984)}]{Nomoto1984}
Nomoto, K. 1984, The Astrophysical Journal, 277, 791

\bibitem[{Nomoto \& Iben(1985)}]{Nomoto1985}
Nomoto, K. \& Iben, I., J. 1985, The Astrophysical Journal, 297, 531

\bibitem[{Nomoto {et~al.}(2007)Nomoto, Saio, Kato, \& Hachisu}]{Nomoto2007}
Nomoto, K., Saio, H., Kato, M., \& Hachisu, I. 2007, The Astrophysical Journal,
  663, 1269

\bibitem[{Paczyński(1971)}]{Paczynski1971}
Paczyński, B. 1971, Acta Astronomica, 21

\bibitem[{Pakmor {et~al.}(2011)Pakmor, Hachinger, R{\"{o}}pke, \&
  Hillebrandt}]{Pakmor2011}
Pakmor, R., Hachinger, S., R{\"{o}}pke, F.~K., \& Hillebrandt, W. 2011,
  Astronomy {\&} Astrophysics, 528, A117

\bibitem[{Pakmor {et~al.}(2010)Pakmor, Kromer, R{\"{o}}pke, Sim, Ruiter, \&
  Hillebrandt}]{Pakmor2010}
Pakmor, R., Kromer, M., R{\"{o}}pke, F.~K., Sim, S.~A., Ruiter, A.~J., \&
  Hillebrandt, W. 2010, Nature, 463, 61

\bibitem[{Pakmor {et~al.}(2012)Pakmor, Kromer, Taubenberger, Sim, R{\"{o}}pke,
  \& Hillebrandt}]{Pakmor2012}
Pakmor, R., Kromer, M., Taubenberger, S., Sim, S.~A., R{\"{o}}pke, F.~K., \&
  Hillebrandt, W. 2012, The Astrophysical Journal, 747, L10

\bibitem[{Paxton {et~al.}(2011)Paxton, Bildsten, Dotter, Herwig, Lesaffre, \&
  Timmes}]{Paxton2011}
Paxton, B., Bildsten, L., Dotter, A., Herwig, F., Lesaffre, P., \& Timmes, F.
  2011, The Astrophysical Journal Supplement Series, 192, 3

\bibitem[{Paxton {et~al.}(2013)Paxton, Cantiello, Arras, Bildsten, Brown,
  Dotter, Mankovich, Montgomery, Stello, Timmes, \& Townsend}]{Paxton2013}
Paxton, B., Cantiello, M., Arras, P., Bildsten, L., Brown, E.~F., Dotter, A.,
  Mankovich, C., Montgomery, M.~H., Stello, D., Timmes, F.~X., \& Townsend, R.
  2013, The Astrophysical Journal Supplement Series, 208, 4

\bibitem[{Paxton {et~al.}(2015)Paxton, Marchant, Schwab, Bauer, Bildsten,
  Cantiello, Dessart, Farmer, Hu, Langer, Townsend, Townsley, \&
  Timmes}]{Paxton2015a}
Paxton, B., Marchant, P., Schwab, J., Bauer, E.~B., Bildsten, L., Cantiello,
  M., Dessart, L., Farmer, R., Hu, H., Langer, N., Townsend, R. H.~D.,
  Townsley, D.~M., \& Timmes, F.~X. 2015, The Astrophysical Journal Supplement
  Series, 220, 15

\bibitem[{Piersanti {et~al.}(2014)Piersanti, Tornambe, \&
  Yungelson}]{Piersanti2014a}
Piersanti, L., Tornambe, A., \& Yungelson, L.~R. 2014, Monthly Notices of the
  Royal Astronomical Society, 445, 3239

\bibitem[{Ritossa {et~al.}(1996)Ritossa, Garcia-Berro, \& Iben}]{Ritossa1996}
Ritossa, C., Garcia-Berro, E., \& Iben, Icko, J. 1996, The Astrophysical
  Journal, 460, 489

\bibitem[{Ritter(1988)}]{Ritter1988}
Ritter, Â. 1988, Astronomy and Astrophysics (ISSN 0004-6361), 202, 93

\bibitem[{Ruiter {et~al.}(2012)Ruiter, Sim, Pakmor, Kromer, Seitenzahl,
  Belczynski, Fink, Herzog, Hillebrandt, Ropke, \& Taubenberger}]{Ruiter2012}
Ruiter, A.~J., Sim, S.~A., Pakmor, R., Kromer, M., Seitenzahl, I.~R.,
  Belczynski, K., Fink, M., Herzog, M., Hillebrandt, W., Ropke, F.~K., \&
  Taubenberger, S. 2012, Monthly Notices of the Royal Astronomical Society,
  429, 1425

\bibitem[{Saio \& Nomoto(1985)}]{Saio1985}
Saio, H. \& Nomoto, K. 1985, Astronomy and Astrophysics (ISSN 0004-6361), 150

\bibitem[{Saio \& Nomoto(1998)}]{Saio1998}
---. 1998, The Astrophysical Journal, 500, 388

\bibitem[{Schwab {et~al.}(2015)Schwab, Quataert, \& Bildsten}]{Schwab2015a}
Schwab, J., Quataert, E., \& Bildsten, L. 2015, Monthly Notices of the Royal
  Astronomical Society, 453, 1910

\bibitem[{Shen \& Bildsten(2007)}]{Shen2007}
Shen, K.~J. \& Bildsten, L. 2007, The Astrophysical Journal, 660, 1444

\bibitem[{Shen {et~al.}(2012)Shen, Bildsten, Kasen, \& Quataert}]{Shen2012}
Shen, K.~J., Bildsten, L., Kasen, D., \& Quataert, E. 2012, The Astrophysical
  Journal, 748, 35

\bibitem[{Shen \& Moore(2014)}]{Shen2014a}
Shen, K.~J. \& Moore, K. 2014, The Astrophysical Journal, 797, 46

\bibitem[{Timmes {et~al.}(1994)Timmes, Woosley, \& Taam}]{Timmes1994}
Timmes, F.~X., Woosley, S.~E., \& Taam, R.~E. 1994, The Astrophysical Journal,
  420, 348

\bibitem[{Townsley \& Bildsten(2004)}]{Townsley2004}
Townsley, D.~M. \& Bildsten, L. 2004, The Astrophysical Journal, 600, 390

\bibitem[{Wang {et~al.}(2015)Wang, Li, Ma, Liu, Cui, \& Han}]{Wang2015}
Wang, B., Li, Y., Ma, X., Liu, D.-D., Cui, X., \& Han, Z. 2015, Astronomy {\&}
  Astrophysics, 584, A37

\bibitem[{Wang {et~al.}(2009)Wang, Meng, Chen, \& Han}]{Wang2009}
Wang, B., Meng, X., Chen, X., \& Han, Z. 2009, Monthly Notices of the Royal
  Astronomical Society, 395, 847

\bibitem[{Wang {et~al.}(2014)Wang, Meng, Liu, Liu, \& Han}]{Wang2014}
Wang, B., Meng, X., Liu, D.-D., Liu, Z.-W., \& Han, Z. 2014, The Astrophysical
  Journal, 794, L28

\bibitem[{Wolf {et~al.}(2013)Wolf, Bildsten, Brooks, \& Paxton}]{Wolf2013}
Wolf, W.~M., Bildsten, L., Brooks, J., \& Paxton, B. 2013, The Astrophysical
  Journal, 777, 136

\bibitem[{Woosley \& Weaver(1986)}]{Woosley1986}
Woosley, S.~E. \& Weaver, T.~A. 1986, IN: Nucleosynthesis and its implications
  on nuclear and particle physics; Proceedings of the NATO Advanced Research
  Workshop (Fifth Moriond Astrophysics Meeting), 145

\bibitem[{Yoon \& Langer(2003)}]{Yoon2003}
Yoon, S.-C. \& Langer, N. 2003, Astronomy and Astrophysics, 412, L53

\end{thebibliography}

\end{document}